\documentclass[12pt,a4paper,english]{article}
\usepackage[T1]{fontenc}
\usepackage[latin1]{inputenc}
\usepackage{amsmath}
\usepackage{amssymb}

\makeatletter

\providecommand{\LyX}{L\kern-.1667em\lower.25em\hbox{Y}\kern-.125emX\@}

 \newcommand{\lyxaddress}[1]{
   \par {\raggedright #1 
   \vspace{1.4em}
   \noindent\par}
 }

\makeatother
\usepackage{babel}
\begin{document}

\title{Limitations of the mean field slave-particle approximations}

\author{E.A. Kochetov$^{1,2}$ and A. Ferraz$^{2}$}

\maketitle

\lyxaddress{$^{1}$Bogoliubov Theoretical Laboratory, Joint Institute for Nuclear
Research, 141980 Dubna, Russia}

\lyxaddress{$^{2}$International Centre for Condensed Matter Physics, Universidade
de Brasília, 70910-900 Brasília- DF, Brazil}

\begin{abstract}
We show that the transformation properties of the mean field slave
boson/fermion order parameters under an action of the global SU(2)
group impose certain restrictions on their applications to describe
the phase diagram of the t-J model. 
\end{abstract}

\section{Introduction}

The mean field (MF) slave-boson/fermion theory is a commonly used
approach to address the t-J model when dealing with spin-charge separation
in the context of a spin liquid, or the resonating valence bond (RVB)
state. Within this scheme a spin-charge separation can be intuitively
implemented representing the electron operator by a product of two
commuting operators that carry separately spin and charge degrees
of freedom. Namely, by introducing the {}``slave boson{}`` (SB)\cite{sb}
one rewrites the on-site electron operator in the form, \begin{equation}
c_{i\sigma }=f_{i\sigma }\, b_{i}^{\dag },\label{eq:1b}\end{equation}
 where $b_{i}$ is a charged spinless (slave) boson operator (holon),
while $f_{i\sigma }$ is a neutral, spin $1/2$ fermion operator (spinon)
satisfying the constraint of no double electron occupancy (NDEO) \begin{equation}
b_{i}^{\dag }b_{i}+\sum _{\sigma =\uparrow ,\downarrow }f_{\sigma i}^{\dag }f_{i\sigma }=1.\label{eq:2b}\end{equation}
 Alternatively, one can also introduce a spinless fermion $f_{i}$
to describe the charge degree of freedom and a {}``spinning{}``
boson $b_{i\sigma }$ to describe the spin degree of freedom. This
is the {}``slave fermion{}`` (SF) approach \cite{sf1,sf2} , \begin{equation}
c_{i\sigma }=b_{i\sigma }f_{i}^{\dag }\label{eq:1f}\end{equation}
 The NDEO constraint now reads \begin{equation}
f_{i}^{\dag }f_{i}+\sum _{\sigma =\uparrow ,\downarrow }b_{\sigma i}^{\dag }b_{i\sigma }=1.\label{eq:2f}\end{equation}

In principle, both the SF and SB theories should produce physically
identical results for the t-J model. However, in the MF approximation
they give very different phase diagrams \cite{fuku,kane,feng}. In
particular, in the SB version the antiferromagnetic (AF) correlation
is absent even for zero doping. Alternatively, in the SF approach,
the ground state is antiferromagnetic for the undoped case and the
long-range order persists until very high doping ($\delta _{c}\sim 0.6$)
\cite{kane}. It is commonly believed that these different results
are due to the fact that in the MF approximation the crucial single
occupancy constraint given by eq.(\ref{eq:2b})/eq.(\ref{eq:2f})
is taken into account only on average. We show however that there
is in fact another important reason for this discrepancy even within
the standard MF approximation. We call attention to the fact that
the SB and SF RVB singlet order parameters (OP) transform in different
ways under a global SU(2) action that leaves the t-J hamiltonian invariant.
While the RVB SB OP $\chi _{ij}^{SB}=<f_{i\uparrow }f_{j\downarrow }-f_{i\downarrow }f_{j\uparrow }>$
is SU(2) invariant and, it is, therefore, more convenient to describe
a phase with unbroken SU(2) symmetry, the SF RVB OP $\chi _{ij}^{SF}=<b_{i\uparrow }b_{j\downarrow }-b_{i\downarrow }b_{j\uparrow }>$breaks
this symmetry explicitly and, therefore, seems more suitable for the
description of the AF ordered state.

\section{General Symmetry Considerations}

Let us start by first discussing the symmetry properties of the t-J
hamiltonian, \begin{equation}
H_{t-J}=-t\sum _{<ij>\sigma }(c_{i\sigma }^{\dag }c_{j\sigma }+H.c.)+J\sum _{<ij>}\left(\textbf S_{i}\textbf S_{j}-\frac{1}{4}n_{i}n_{j}\right),\label{eq:tj}\end{equation}
 where $\textbf S_{i}=c_{i}^{\dag }\sigma c_{i}/2$ - electron spin
operators with $\sigma $ as Pauli matrices, and $n_{i}=\sum _{\sigma }c_{i\sigma }^{\dag }c_{i\sigma }$
is the electron number operator. The hamiltonian (\ref{eq:tj}) is
defined in a restricted Hilbert space without double electron occupancy.

It is clear, that the total number of the electrons, $N=\sum _{i}n_{i}$
is conserved, which results in the global U(1) symmetry of eq.(\ref{eq:tj}).
Besides, the spin operators $\textbf S=\sum _{i}\textbf S_{i}$ generate
global SU(2) rotations of the electron operators $(c_{\uparrow },c_{\downarrow })$
which transform as SU(2) doublet, \begin{eqnarray}
\left(\begin{matrix} c_{i\uparrow }\\
 c_{i\downarrow }\end{matrix}\right)\rightarrow \left(\begin{matrix} c_{i\uparrow }\\
 c_{i\downarrow }\end{matrix}\right)^{\prime }=\left(\begin{matrix} u & v\\
 -\bar{v} & \bar{u}\end{matrix}\right)\left(\begin{matrix} c_{i\uparrow }\\
 c_{i\downarrow }\end{matrix}\right),\quad \left(\begin{matrix} u & v\\
 -\bar{v} & \bar{u}\end{matrix}\right)\in SU(2), &  & \label{eq:transf}
\end{eqnarray}
 leaving again the hamiltonian (\ref{eq:tj}) invariant. Note that
the SU(2) group parameters $u$ and $v$ are taken to be site-independent.Thus
the t-J hamiltonian (\ref{eq:tj}) possesses the global $U(1)\times SU(2)$
symmetry.

Within the MF approximation the spin liquid phase of the t-J model
is believed to be adequately described by the globally SU(2) invariant
RVB electron spin singlet OP $\chi _{ij}\equiv <c_{i\uparrow }c_{j\downarrow }-c_{i\downarrow }c_{j\uparrow }>$\cite{bask}.
It however breaks the U(1) global symmetry related to the conservation
of the total number of the electrons. In the slave-particle representations
the RVB OP takes on the following representations, \[
\chi _{ij}=<b_{i}^{\dag }b_{j}^{\dag }><f_{i\uparrow }f_{j\downarrow }-f_{i\downarrow }f_{j\uparrow }>\]
 or \[
\chi _{ij}=<f_{i}^{\dag }f_{j}^{\dag }><b_{i\uparrow }b_{j\downarrow }-b_{i\downarrow }b_{j\uparrow }>.\]
 Although both decompositions of $\chi _{ij}$ are SU(2) invariant
their single constituents in general need not be so. This is because
there is an additional U(1) local gauge invariance under the transformation
$f_{i}\rightarrow f_{i}\, e^{i\vartheta _{i}},\, b_{i}\rightarrow b_{i}\, e^{-i\vartheta _{i}}$
that leaves eqs.(\ref{eq:1b},\ref{eq:1f}) intact. To appropriately
reduce the number of degrees of freedom , one should {}``gauge-fix{}``
$\vartheta _{i}$. The important point is that the gauge fixing must
be SU(2) invariant. In other words, the gauge fixing must be compatible
with the SU(2) invariance of the RVB OP $\chi _{ij}$. As we shall
see, this imposes some restrictions on the transformation properties
of the $f$ and $b$ fields.

\section{Slave Fermion Representation}

Let us, first, consider the SF case. It will be more convenient to
deal with the SF path-integral representation of the t-J partition
function. Within that representation the classical counterparts of
eqs.(\ref{eq:1f}) and (\ref{eq:2f}) read \begin{equation}
c_{i\sigma }=b_{i\sigma }\bar{f}_{i},\label{eq:3f}\end{equation}
\begin{equation}
\bar{f}_{i}f_{i}+\sum _{\sigma =\uparrow ,\downarrow }\bar{b}_{\sigma i}b_{i\sigma }=1,\label{eq:4f}\end{equation}
 respectively, where now $c_{i\sigma }$ and $f_{i}$ are complex
Grassmann parameters, whereas $b_{i\sigma }$ stands for complex c-numbers.The
OP´s are now understood to be the path-integral everages, e.g., \begin{equation}
<b_{i\uparrow }b_{j\downarrow }-b_{i\downarrow }b_{j\uparrow }>=\int D\mu (b_{i\uparrow }b_{j\downarrow }-b_{i\downarrow }b_{j\uparrow })e^{S_{t-J}^{SF}(f,b_{\uparrow },b_{\downarrow })}/\int D\mu e^{S_{t-J}^{SF}(f,b_{\uparrow },b_{\downarrow })},\label{eq:opsf}\end{equation}
 where $S_{t-J}^{SF}(f,b_{\uparrow },b_{\downarrow )})$ is the t-J
action in the SF representation (\ref{eq:3f}).

It is clearly seen that eq.(\ref{eq:3f}) increases the number of
degrees of freedom by two. The constraint (\ref{eq:4f}) takes care
of one of them, and the extra one must be dealt with by the fixing
of the U(1) local gauge. This is achieved by fixing the phase of one
of the bosonic fields, by requiring, e.g., that $\arg b_{i\downarrow }=0$.
In other words, to fix the gauge, we impose the condition \begin{equation}
\arg b_{i\downarrow }=\frac{1}{2i}\log \frac{b_{i\downarrow }}{\bar{b}_{i\downarrow }}=0.\label{eq:arg1}\end{equation}

Let us first assume that the $b_{i\sigma }$ fields transform in a
linear spinor representation of SU(2) just as true fermionic amplitudes:
\[
b_{i\uparrow }^{\prime }=ub_{i\uparrow }+vb_{i\downarrow },\]
\begin{equation}
b_{i\downarrow }^{\prime }=\bar{u}b_{i\downarrow }-\bar{v}b_{i\uparrow }\label{eq:b}\end{equation}
 If this is the case, the slave fermion $f_{i}$ should be a SU(2)
scalar. However calculating the phase of the transformed operators
gives \[
\arg b_{i\downarrow }^{\prime }=\frac{1}{2i}\log \frac{b_{i\downarrow }^{\prime }}{\bar{b}_{i\downarrow }^{\prime }}=\frac{1}{2i}\log \frac{-\bar{v}b_{i\uparrow }+\bar{u}b_{i\downarrow }}{-v\bar{b}_{i\uparrow }+u\bar{b}_{i\downarrow }}=\frac{1}{2i}\log \frac{-\bar{v}z_{i}+\bar{u}}{-v\bar{z}_{i}+u}\neq 0,z_{i}\equiv b_{i\uparrow }/b_{i\downarrow }.\]
 This tells us that eq.(\ref{eq:arg1}) is not truly SU(2) covariant.
Nevertheless, the covariance can be restored if we multiply eq.(\ref{eq:b})
by an appropriate phase factor: \[
b_{i\uparrow }^{\prime }=e^{i\psi _{i}}(ub_{i\uparrow }+vb_{i\downarrow }),\]
\begin{eqnarray}
b_{i\downarrow }^{\prime }=e^{i\psi _{i}}(\bar{u}b_{i\downarrow }-\bar{v}b_{i\uparrow }), &  & \label{eq:okb}
\end{eqnarray}
 where \begin{equation}
i\psi _{i}=\frac{1}{2}\log \frac{-v\bar{z}_{i}+u}{-\bar{v}z_{i}+\bar{u}}.\label{eq:psi}\end{equation}
 In this way we can guarantee that $\arg b_{i\downarrow }^{\prime }=0.$
The same kind of phase factor shows up in the transformation law of
the SU(2) covariant Kaehler potential $K=s\log (1+|z|^{2})$ for a
spin $s$. In fact, under SU(2) rotations of the two-sphere $\textrm{S}^{\textrm{2}}$,
or, equivalently, of the projective space $CP^{1}$, one gets \[
K\rightarrow K+\varphi +\bar{\varphi },\quad \varphi =-s\log (-\bar{v}z+\bar{u}),\]
 so that $i\psi =\varphi -\bar{\varphi }$ at $s=1/2$. Equation (\ref{eq:2f})
defines a supersphere $CP^{1|1}$ (see Appendix) whose body\cite{body}
coincides with the $CP^{1}$ manifold. Since $CP^{1}$ is a compact
manifold, $SU(2)$ acts on it in a nonlinear way. For this reason,
the function $\psi $ is a natural ingredient in the SU(2) transformation
law for the SF fields.

Since the true electron operators $c_{i\sigma }$ are by definition
transformed according to eq.(\ref{eq:transf}) we conclude that the
slave fermion must transform as \begin{eqnarray}
f_{i}\rightarrow f_{i}^{\prime }=e^{-i\psi _{i}}f_{i}. &  & \label{eq:okf}
\end{eqnarray}
 Despite the explicit site dependence of $\psi _{i}$ through the
$z_{i}$ field, eqs.(\ref{eq:okb},\ref{eq:okf}) represent \emph{global}
SU(2) transformations (the group parameters $u$ and $v$ are site
- independent).This transformation has also nothing to do with the
above discussed local U(1) gauge invariance of the t-J model in the
slave-particle representation. In fact we have already taken care
of that gauge freedom by imposing the condition (\ref{eq:arg1}).

As is shown in the Appendix our somewhat heuristic argumentation that
lead to (\ref{eq:okb},\ref{eq:okf}) can be made rigorous by employing
the $su(2|1)$ superalgebra representation of the Hubbard operators.
Such a representation follows if we explicitly resolve the constraint
of no double occupancy (\ref{eq:4f}) which is basically an equation
of the $SU(2|1)$ homogeneous supersphere embedded into a flat superspace.
The spin group SU(2), being a subgroup of $SU(2|1)$, acts on a supersphere
homogeneously and in a nonlinear way, which reasserts itself in the
highly nonlinear transformation laws for the $f$ and $b_{\sigma }$
fields under the SU(2) action.

Since both the SF action and the measure factor in eq.(\ref{eq:opsf})
are SU(2) invariant, this means that, under (\ref{eq:okb},\ref{eq:okf}),
the SF RVB OP´s are \emph{not} SU(2) invariant. They transform simply
as \begin{eqnarray}
<b_{i\uparrow }b_{j\downarrow }-b_{i\downarrow }b_{j\uparrow }> & \rightarrow  & e^{i(\psi _{i}+\psi _{j})}<b_{i\uparrow }b_{j\downarrow }-b_{i\downarrow }b_{j\uparrow }>,\\
<f_{i}^{\dag }f_{j}^{\dag }> & \rightarrow  & e^{-i(\psi _{i}+\psi _{j})}<f_{i}^{\dag }f_{j}^{\dag }>.\label{eq:finsf}\nonumber 
\end{eqnarray}
 As a result this naturally explains why the use of the SF OP´s is
more appropriate for the description a phase with a broken SU(2) magnetic
symmetry and may produce quite unreliable results for the doping regions
which are not magnetically ordered. This has already been implicitly
confirmed by direct calculations in the SF MF approximation\cite{kane}.

\section{Slave Boson Representation}

We turn now to the SB case. Within the SB path-integral representation
of the t-J partition function we get the operator classical counterparts
\begin{equation}
c_{i\sigma }=f_{i\sigma }\bar{b}_{i},\label{eq:3b}\end{equation}
\begin{equation}
\bar{b}_{i}b_{i}+\sum _{\sigma =\uparrow ,\downarrow }\bar{f}_{\sigma i}f_{i\sigma }=1,\label{eq:4b}\end{equation}
 where now $c_{i\sigma }$ and $f_{i\sigma }$ are complex Grassmann
parameters, and the $b_{i}$´s stand for complex c-numbers. We can
now fix the local U(1) gauge by choosing $\arg b_{i}=0$. Since Grassmann
parameters are not $c$-valued numbers, we are not able to fix the
phase of the $f_{\sigma }$ field, by demanding, e.g.,that $\log \frac{f_{\downarrow }}{\bar{f_{\downarrow }}}=0$.
This expression is just meaningless for Grassmann variables..

It can easily be checked that the SU(2) transformations lead to \begin{eqnarray}
f_{i\uparrow }^{\prime } & = & uf_{i\uparrow }+vf_{i\downarrow },\\
f_{i\downarrow }^{\prime } & = & \bar{u}f_{i\downarrow }-\bar{v}f_{i\uparrow }\nonumber \\
b_{i}^{\prime } & = & b_{i}\label{eq:fin}\nonumber 
\end{eqnarray}
 which are compatible with the gauge fixing condition, $\arg b_{i}=0$.
Therefore, the SB RVB OPs $<f_{i\uparrow }f_{j\downarrow }-f_{i\downarrow }f_{j\uparrow }>$
as well as $<b_{i}^{\dag }b_{j}^{\dag }>$ are SU(2) invariant and
are more suitable to the description of the doping range not associated
with magnetic ordering, i.e., the superconducting phase\cite{fuku}.

\section{Conclusion}

To conclude, mathematically, the distinctions in the transformation
properties between SF and SB amplitudes can be attributed to the fact
that eq.(\ref{eq:4f}) defines a supermanifold, $CP^{1|1}$ that has
a \emph{compact} body manifold $CP^{1}$. In contrast, eq.(\ref{eq:4b})
defines a supermanifold $CP^{0|2}$ which is essentially fermionic
and contains no compact body manifold. Our results explain quite naturally
why the SF mean field approximations produce qualitatively good results
for magnetically ordered state whereas the SB representation is more
appropriate to represent the superconducting state at larger dopings.

\emph{Ackowledgements:} E.A.K wants to acknowledge the hospitality
of the ICCMP´s staff and the financial support received from CAPES
- Brazil.

\section*{Appendix}

In this Appendix we derive rigorously eqs.(\ref{eq:okb},\ref{eq:okf}).

First, we show that constraint of no double occupancy (\ref{eq:4f})
is explicitly resolved in terms of the $su(2|1)$ path-integral representation
used in Refs.\cite{k1}. We start with the path-integral SF representation
of the t-J partition function (\ref{eq:opsf}). Basic ingredients
that enter the SF path-integral action are the classical symbols of
the SF Hubbard operators $X$. Let $X_{\lambda \lambda `},\, \lambda =1,2,3$
be a $3\times 3$ matrix of the Hubbard operator $X$. Consider a
complex composite vector $d^{t}=(b_{\uparrow },b_{\downarrow },f)^{t}$.
Then, the SF representation reads $X^{cl}=\sum _{\lambda }\bar{d}_{\lambda }X_{\lambda \lambda `}d_{\lambda `},$
where \[
\sum _{\lambda }\bar{d}_{\lambda }d_{\lambda }=\bar{b}_{\uparrow }b_{\uparrow }+\bar{b}_{\downarrow }b_{\downarrow }+\bar{f}f=1\]
 at every lattice site. Let us now make a change of variables that
explicitly resolves this constraint, \[
b_{\uparrow }=\frac{z}{\sqrt{1+|z|^{2}+\bar{\xi }\xi }},\quad b_{\downarrow }=\frac{1}{\sqrt{1+|z|^{2}+\bar{\xi }\xi }},\]
\begin{equation}
f=\frac{\xi }{\sqrt{1+|z|^{2}+\bar{\xi }\xi }}.\label{eq:constr}\end{equation}
 Geometrically, the set $(z,\xi )$ appears as local (projected) coordinates
of a point on the supersphere $CP^{1|1}$ defined by eq.(\ref{eq:4f}).They
are related to the homogeneous (defined up to a scaling factor) coordinates
by $z=b_{\uparrow }/b_{\downarrow },\xi =f/b_{\downarrow },\, b_{\downarrow }\neq 0.$
Note that according to our choice, $\arg b_{\downarrow }=0.$

In terms of the local coordinates, SU(2) acts on a supersphere by
the linear fractional transformations, \begin{equation}
z\rightarrow z^{\prime }=\frac{uz+v}{-\bar{v}z+\bar{u}},\quad \xi \rightarrow \xi ^{\prime }=\frac{\xi }{-\bar{v}z+\bar{u}},\quad \left(\begin{matrix} u & v\\
 -\bar{v} & \bar{u}\end{matrix}\right)\in SU(2),\label{eq:2}\end{equation}
 Substituting this into eq.(\ref{eq:constr}) results in eqs.(\ref{eq:okb},\ref{eq:okf}).

\end{document}